\documentclass[useAMS,usegraphicx]{mn2e}

\newcommand{\et}{\sl et al \rm}

\title[Blue, green, and red bumps in AGN]{Blue, green, and red bumps in AGN}

\author[A. Lawrence]{A.Lawrence$^{1}$ \\
$^{1}$ Institute for Astronomy, University of Edinburgh, Royal Observatory,
Blackford Hill, Edinburgh EH9 3HJ }

\begin{document}

\date{Accepted 2005 July 4th. Received 2005 April 18 ; in original form
2004 July
11}

\pagerange{\pageref{firstpage}--\pageref{lastpage}} \pubyear{2004}

\maketitle

\label{firstpage}

%++++++++++++++ ABSTRACT +++++++++++++++++++++++++++++
\begin{abstract}

I show that the summed spectral energy distribution (SED) $L(\nu)$ of any
 extended blackbody
radiator will scale in a predictable way if all parts of the body change in
 temperature by
the same factor $X$, such that $ L^\prime(\nu) = X^3L(\nu/X) $. This should for
 example
apply to accretion disks around black holes, where $X$ is relative accretion
 rate, or external
heating rate, but will not apply to changes in black hole mass. I summarise
 evidence that
AGN optical-UV SEDs become progressively redder with decreasing luminosity, and
 show that
the trend in colour versus luminosity shown by Mushotzky and Wandel (1989) is
 matched
extremely well by taking a template high-luminosity SED and scaling it in the
 manner
described above. This agreement is striking because it involves {\em no
adjustable parameters}. The agreement breaks down at low luminosities because of
 stellar
contamination and reddening. I then consider the colour changes of an
individual AGN (NGC~5548) during luminosity changes, which according to the
 popular X-ray
reprocessing model, should follow the scaling law well. However the observed
 changes
are clearly {\em not} consistent with the simple scaling prediction. Instead,
 these colour
changes are quite well explained by the mixing of a constant red component and
 variable
blue component. Overall, there is then strong support for the ideas
(i) that AGN optical-UV SEDs arise from accretion discs, (ii) that accretion
 rate plays a
significant role in the very large range
of luminosity seen in AGN, and (ii) that the
inner regions of AGN vary independently of the outer accretion disc.

\end{abstract}

\begin{keywords}

galaxies:active -- galaxies:quasars:general -- accretion -- accretion discs.

\end{keywords}

%++++++++++ BODY OF PAPER ++++++++++++++++++++++++++

\section{Introduction}  \label{intro}

The spectral energy distribution (SED) of a typical quasar is dominated
by a broad optical-UV hump  normally identified
with the emission
expected from an accretion disc surrounding a massive black hole, and often
referred to as  ``Big Blue Bump''.
The effective power law index (i.e. $\alpha= - d\log S_\nu / d\log \nu$)
is $\alpha\sim 0$ in the optical region,
steepening to $\alpha\sim 1$ in the UV (see for example the  compilation of
 Elvis \et
(1994)). The composite spectrum of high
redshift quasars compiled by Zheng \et\ (1997) shows that the spectral
index steepens further to $\alpha\sim 2$ in the far UV, and this seems
consistent with extrapolation to the soft X-ray excess visible in most
quasars.

%%%%%%%%%%%%%%%%%%%%%%%%%%%%%%%%%%%%%%%%%%%%%%%%%%%%%%%
% FIGURE-1 : SED COMPARISONS
%
\begin{figure}

\includegraphics[width=60mm,angle=270]{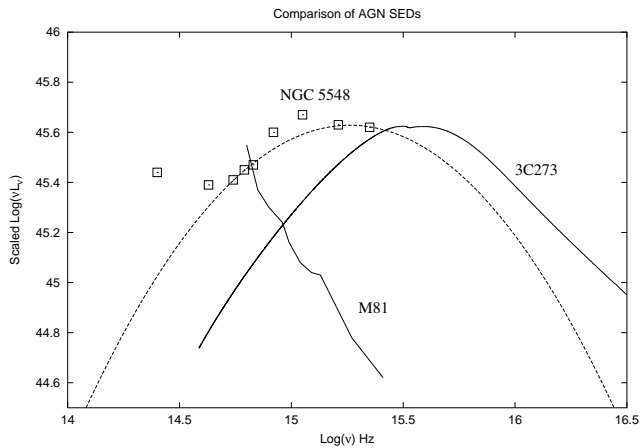}

\caption{Comparison of AGN spectral energy distributions.
For 3C273, the line shown is a smooth model curve taken from Kriss et al 1999
which fits their optical-UV-soft X-ray data but which excludes the ``small blue
 bump''.
For NGC~5548 the data points are from Rokaki and Magnan 1992 and Ward et al
 1987, and
the dashed line is an illustrative parabola. For M81 the line is based on data
 in
Ho, Filippenko and Sargent 1996. }

\end{figure}

%%%%%%%%%%%%%%%%%%%%%%%%%%%%%%%%%%%%%%%%%%%%%%%%%%%%%%%

Not all Active Galactic Nuclei (AGN) are the same however. Figure 1
compares the SEDs of three AGN. The first is the luminous quasar 3C273,
which shows the characteristic UV Bump as described above. The curve
shown is actually a model fit to optical, UV, and soft X-ray data by
Kriss \et (1999), which excludes the line emission and ``small blue bump''
composed of Balmer continuum and quasi-continuum of FeII lines.
(The model curve is that in Fig. 8 of Kriss \et and was kindly
provided by G.Kriss).
The second AGN is the Seyfert galaxy NGC 5548. This object is both
variable and includes substantial starlight emission in the low state.
The data points shown in Figure 1 represent the optical-UV mean of Rokaki
and Magnan (1992) together with the near-IR data of Ward \et (1987), after
 correcting
for starlight emission following the data of Romanishin \et (1995).
The curve shown is a log-quadratic hand-fitted to the data points for
illustrative purposes. Note that the U-band and 2670$\rm\rm\AA$ points certainly
contain substantial ``small blue bump'' emission. I have not quantitatively
modelled this; the curve shown is only a qualitative indication of the
likely underlying continuum. The result is that the NGC 5548 Bump seems
similar to that of 3C273 but shifted to lower frequency -
more of a ``Big Green Bump'' than a ``Big Blue Bump''. Finally, the third
curve shown is the SED of the dwarf Seyfert nucleus in M81. The data has
been adapated from the tables and figures of Ho, Filippenko and Sargent (1996),
who applied a de-reddening using the ratios of observed narrow emission lines.
The optical emission is from an unresolved point source in HST imaging,
and so seems to be truly nuclear, with negligible starlight. This SED,
with a UV slope of $\alpha\sim 2$, is quite different from that of luminous
quasars, and indeed Ho, Filippenko and Sargent stress that dwarf Seyferts
may not be completely analogous to luminous quasars. However, the optical
slope seen in M81 is strikingly similar to that seen in the far-UV for
luminous quasars, leading one to speculate that it too has a Bump, but
shifted to much lower frequency - a ``Big Red Bump'' ?

%%%%%%%%%%%%%%%%%%%%%%%%%%%%%%%%%%%%%%%%%%%%%%%%%%%%%%%
% FIGURE-2 : ALPHA-L EFFECT : MW DATA
%
\begin{figure*}

\includegraphics[width=100mm,angle=270]{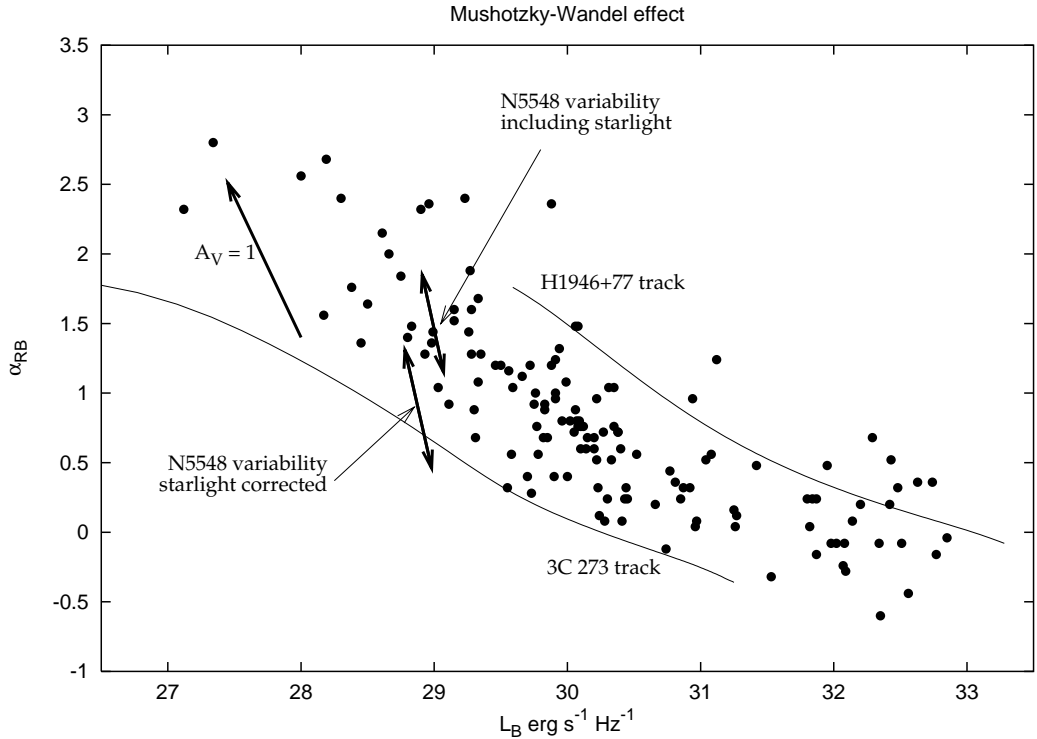}

\caption{Correlation between narrow band optical luminosity, measured at
$4200 \rm\AA$, and spectral index between $4200 \rm\AA$ and $7500 \rm\AA$. The
 data points are
 taken from Mushotzky and Wandel (1989). The two curves are the tracks predicted
 by taking
SEDs of 3C273 and H1946+786 respectively, and shifting them homologously as
 expected for
a multi-temperature black body, as explained in the text. Also shown is a
 reddening vector
assuming standard Galactic reddening, and the high and low states of NGC~5548,
 both before
and after starlight subtraction (see text).}

\end{figure*}

%
%%%%%%%%%%%%%%%%%%%%%%%%%%%%%%%%%%%%%%%%%%%%%%%%%%%%%%%

That the continuum shape of AGN varies slowly and systematically with
luminosity has been argued several times by previous authors. Figure 2
shows data collated by  Mushotsky and Wandel (1989), demonstrating that
optical slope varies with optical luminosity, ranging from $\alpha_{opt}\sim 0$
at the highest luminosities, through $\alpha_{opt}\sim 1$ in the
Seyfert galaxy range, to $\alpha_{opt}\sim 2$ for very low luminosity AGN.
(Starlight contamination and reddening is undoubtedly important in this figure;
I consider this in more detail in Section 3.) A similar effect was shown
in the RIXOS sample by Puchnarewicz \et (1996) and an analogous effect
for UV slope  was shown by Zheng and Malkan (1993).

There are also arguments against the whole AGN SED shifting bodily
with luminosity. From an observational point of view, the fact that
broad emission line ratios change little with luminosity means
that the FUV/soft-X-ray spectrum cannot change much. From a theoretical
point of view, we might expect some scaling of accretion disc properties
with luminosity, but this is least likely to apply shortward of the SED peak,
as I explain in Section 3. Nontheless the evidence concerning the systematic
trend of optical colour with luminosity is very clear, and scaling
of accretion disc properties seems the best possibility to explore.
(Other possible explanations of the colour-luminosity trend are
considered in Section 5.)

The hypothesis I explore then in this paper is that this luminosity
dependence of AGN is due to a quasi-universal SED which shifts to
increasing frequency with increasing luminosity, and which results from
the expected behaviour of accretion discs. Such shifting is qualitatively
what one expects of a blackbody emitter changing temperature, but of
course accretion discs are not single temperature objects, and their
expected SEDs are still somewhat controversial, so this needs a little thought.
In section 2 I derive a scaling law for the behaviour of multi-temperature
blackbodies, and discuss how this applies to accretion discs. In section 3
I compare the predicted behaviour with that seen in the AGN data. In section 4
I ask whether the same behaviour applies to individual AGN during variability.
Finally in section 5 I discuss the general implications for AGN models.

\section{Colour-luminosity scaling for black-body emitters}
\label{scaling}

\subsection{Black body scaling with temperature} \label{BB}

As a blackbody changes temperature, it shifts in frequency and also
scales in brightness. This shifting
can be expressed as a simple scaling law that is very useful when we
come to consider objects that emit
the sum of many blackbodies. Consider the blackbody function
\[
R(\nu, T) = \frac{2\pi h \nu^3}{c^2(e^{h\nu/kT}) - 1}
\]
Suppose initially the temperature is $T_0$ and then changes by a
factor $X$ so that $T=XT_0$. Then
\[
R(\nu, T) = \frac{2\pi h \nu^3}{c^2} /
\left[  exp\left(\frac{h}{k} \frac{\nu}{X} \frac{1}{T_0} \right) - 1  \right]
\]
Let $\nu/X = \nu_0$ ; then
\[
R(\nu, T) = \frac{2\pi h }{c^2}  \nu_0^3X^3 /
\left[  exp\left( \frac{h\nu_0}{kT_0}\right) - 1  \right]
\]
i.e. $R(\nu,T) = X^3 R(\nu_0, T_0) $ or
\begin{equation}
R_T(\nu) = X^3R_{T_0}(\nu_0 = \nu/X)
\end{equation}

Thus blackbody SEDs show a kind of homologous scaling with temperature.
The equation above can be interpreted as follows.
If you know (numerically) the blackbody function at some
temperature $T_0$, then the equation gives a recipe for creating the
function at any other temperature $T$
step by step at each frequency.

\subsection{Multi-temperature black bodies} \label{multi-BB}

We can show that under certain simple conditions, an object emitting
as the sum of blackbodies follows the
same homologous scaling law as derived above for single blackbodies.
Consider two distinct regions. The
first has
area $A_1$ and is at temperature $T_1$.
At frequency $\nu$ it will have an SED given by $L_1(\nu) = A_1
R_{T_1}(\nu)$.
Now suppose that its temperature changes by a factor $X_1$ to become $T^\prime_1
 = X_1
T_1$. Following equation (1) the new SED will be given
by $L^\prime_1(\nu) = A_1 X_1^3 R_{T_1}(\nu/X_1)$.

Likewise area $A_2$ at temperature $T_2$ has SED $L_2(\nu) = A_2 R_{T_2}(\nu)$,
 but
when its temperature
changes by factor $X_2$ then its SED
becomes $L^\prime_2(\nu) = A_2 X_2^3 R_{T_2}(\nu/X_2)$. Now we
consider the summmed emission. Before the temperature change this is
\[
L(\nu) = A_1 R_{T_1}(\nu) + A_2 R_{T_2}(\nu)
\]
and after the temperature change it is
\[
L^\prime(\nu) = A_1 X_1^3 R_{T_1}(\nu/X_1) + A_2 X_2^3 R_{T_2}(\nu/X_2)
\]
Now if $X_1=X_2=X$ then
\[
L^\prime(\nu) = X^3 \left( A_1 R_{T_1}(\nu/X)
+ A_2 R_{T_2}(\nu/X) \right)
\]
and so
\begin{equation}
L^\prime(\nu) = X^3L(\nu/X)
\end{equation}

By repeated addition one can see that our scaling law applies to any
sum of blackbodies,
with many
sub-areas at various temperatures, as long as each sub-area
changes temperature by the {\em same factor X}.
The beauty of this is that one does not need an analytic expression for
 $L(\nu)$, or even
an understanding
of its origin - only a reasonably fine
grained measurement of it. Given this ``template'' SED, one can numerically
 construct any
other SED given
a uniform temperature change by a factor $X$.  But is this restricted scenario
 of uniform
temperature
change relevant to the accretion disc problem ?

\subsection{Expected behaviour of accretion discs} \label{discs}

Sophisticated accretion disc models have been developed and applied to AGN for
 some years (e.g. Sun and Malkan 1986). My aim here however is to draw out
the basic points to see whether or not homologous scaling will apply.

To a good approximation, an accretion disc is the sum of many blackbodies. The
 temperature
at radius $R$ will be of the order of that given by the local release of binding
 energy
\[
T_{BE}(R) = \left( \frac{GM_H \dot{m}}{8\pi R^3 \sigma} \right)^{1/4}
\]
where $M_H$ is the central black hole mass, $\dot{m}$ is the accretion rate and
 $\sigma$ is
the
Stefan-Boltzmann constant. The correct $T(R)$ formula depends on the physical
 model - for
example assuming Newtonian dynamics and zero stress at an inner boundary $R_*$
gives $T^4 = T_{BE}^4\left(
1- \left( R_*/R\right)^{1/2}\right) $
(Frank, King and Raine 2002 p.90, Krolik 1999, p.148). However, the
scaling with $M_H$ and $\dot{m}$ is just the same.
If for example $\dot{m}$ changes by a factor $Y$, the temperature
changes by the same factor $X=Y^{1/4}$ at all radii, and we should expect to
 witness
homologous scaling of
accretion disc SEDs under changes of accretion rate.

At first, it seems that the same effect should apply to changes
in black hole mass, but not when one considers the boundary condition, which is
 that the {\em
inner radius} of
the disc will change with $M_H$. When changing  from a small black hole to a
 large one, it
is not the
case that all annuli change temperature by the same factor, as the innermost
 radii cease
radiating at all.
The maximum temperature will be lower, and hence the peak frequency lower, and
 in this
sense the SED from
a large black hole looks ``cooler'' than the radiation from a small black hole.
It is still the case that at
each radius, the temperature gets {\em larger} with increasing black hole mass.
 One might
therefore hope that if one was looking at wavelengths where the radiation is not
 dominated
by emission from the inner radii, then the homologous scaling would still work.
 However a
little numerical experimentation shows that this is not the case.
 Figure 3 shows some toy models of
accretion disc emission using the simple $T_{BE}$ formula. This confirms that
 the homologous
scaling works quite accurately for changes in accretion rate, but that changes
 in black hole
mass produce an effect very different from homologous scaling. The manner in
 which mass changes
alter the SED will depend on the actual $T(R)$ formula, and so is model
 dependent, but the
accretion rate scaling should be generally applicable. Some similar
experimentation with full relativistic modelling of accretaion disks (Malkan
1991) shows a similar result - a homologous scaling with accretion rate, but not
with black hole mass.

%%%%%%%%%%%%%%%%%%%%%%%%%%%%%%%%%%%%%%%%%%%%%%%%%%%%%%%
% FIGURE-3 : ACCRETION  DISC MODELS
%
\begin{figure}
\includegraphics[width=60mm, angle=270]{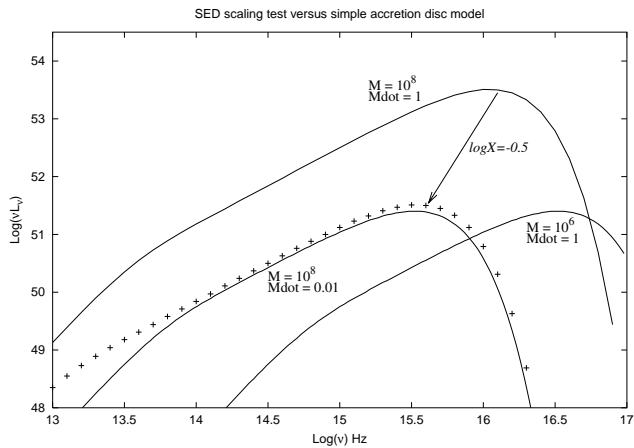}
\caption{Comparison of simple accretion disc models with the
homologous scaling prediction. The model assumes temperature as a function of
radius given by the local binding energy, inner radius at $3R_S$, and outer
 radius
at $1000 R_S$. The lower two models differ from the upper model
by a factor of 100 in accretion rate and black hole mass respectively. The
 discrete points
show the result of applying the homologous shift formula with $X=100^{1/4}$.}
\end{figure}

%
%%%%%%%%%%%%%%%%%%%%%%%%%%%%%%%%%%%%%%%%%%%%%%%%%%%%%%%

The above analysis applies to gravitationally heated discs, but a very similar
argument applies to passive discs with an external heating source. For example,
 a flat
passive disc which is
heated by a point source of luminosity $L_*$ at a height $H$ above the disc will
 have a
temperature
profile given by
\[
T(R) = \left(   \frac{H}{8\pi\sigma} . \frac{L_*}{R^3}      \right) ^{1/4}
\]
For changes in heating rate $L_*$, the temperature will change by the same ratio
  at all radii, and
so one expects
homologous scaling.

So what are the limits on the applicability of homologous scaling to accretion
 discs in AGN
? The first
issue is
the degree to which the radiation from an accretion disc will be blackbody like.
 Through
most of the SED,
this will be a good approximation - as with stars, disk atmosphere effects will
 be a second
order affair.
However at high frequencies, over the peak of the SED, this may not be the case,
 with
Compton scattering and other  effects
producing an extended tail. The second issue is that most simple models assume
 some kind of
heating and
radiation on the spot. This is most likely to break down in the inner radii.
 Overall the
expectation that
accretion discs should show homologous scaling is quite good, and most likely to
 apply at
wavelengths
longward of the Big Blue Bump peak, i.e. in the optical and near ultraviolet,
and not in the far ultraviolet or soft X-ray. The most
likely parameters causing scaling effects will be accretion rate or heating
 rate. Varying
black hole mass will not produce homologous scaling.

\section{Comparison with data}
\label{comparison}

\subsection{The colour-luminosity relation}
\label{comp-colour-lum}

From our scaling law, we construct a recipe for calculating the spectral index
 between two
fixed
wavelengths as a function of the monochromatic luminosity at one of those
 wavelengths. The
starting point
needed is a numerical estimate of the SED of one particular actual quasar,
 $L_0(\nu)$. Then,
using equation (2),
other objects
will have SED $L(\nu)=X^3 L_0(\nu/X)$ where X is some unknown factor which
 controls
temperature at each
radius. Consider two fixed frequencies $\nu_1$ and $\nu_2$. The observed flux
 ratio as a
function of $X$ will be
\begin{equation}
R_X = \frac{L(\nu_1)}{L(\nu_2)} = \frac{L_0(\nu_1/X)}{L_0(\nu_2/X)}
\end{equation}

So starting with the observed $L_0(\nu)$ and varying $X$ one generates a
 sequence of points
($R_X,
L_X(\nu_1)$) which can be plotted on the colour-luminosity plane.

I chose two observed template SEDs for high luminosity quasars. The reason for
 choosing
two is that there
is some dispersion in the observed SEDs of AGN at any given luminosity, and one
 wants to
bracket the
observed range. The first template SED is that of 3C273, using the model fit of
 Kriss
\et (1999) as shown in Figure 1.
This does not
mean we are testing the specific model assumptions of Kriss \et\ - the curve is
 simply a
convenient smooth
curve that approximates well to the observed data. The second template SED is
 that of
H1946+786 given in
Kuhn \et\ (1995). As this quasar is at high redshift, there is some uncertainty
 about the
high frequency
correction to be applied. Kuhn \et\ show a range of such corrections.  We choose
 the curve
which gives a
high frequency SED similar to the average found by Zheng \et\ (1997). (This
 makes little
difference in the
optical wavelength range). The curve was digitised from the plot in the Kuhn \et
 paper.

As we start at high luminosity, constructing a locus to lower luminosity
 involves values
$X<1$. Fig.2
shows the loci starting from the two templates at $X=1$ through to values of
 $X=0.01$.  The
two loci each
re-produce the general trend of colour with luminosity, and even the curvature
 in the
relation. Between
the two templates, the range of SEDs is well covered. Although the agreement is
 not
perfect, it is quite
remarkable in one important respect : given a particular starting template, the
 locus has
{\em no free
parameters whatsoever}. It is a strict and simple prediction on the assumption
 that the SED
is made of
many blackbodies, and that we are well away from boundary conditions.

The fit seems least good at very low luminosities, but this may be expected for
 a number of
reasons.
First, the Bump is shifted so far to low frequencies that even at optical
 wavelengths we
are looking over
the peak at radiation probably produced in the inner regions. Second, the
colours may be affected by reddening. Indeed, reddening could of course broaden
 the locus
at all
luminosities. Fig. 2 shows a standard reddening vector to indicate this effect.
Finally, even though Mushotzky and Wandel (1989) used small slit observations,
 there is
very likely to be residual starlight affecting these data. For the Seyfert
 galaxy NGC 5548,
the stellar contribution has been well studied (Romanishin \et 1995). Figure 2
 displays NGC
5548 data with and without the starlight correction, showing that the effect is
 about the
right size to produce the divergence of the bulk of the data points from the
 locus
prediction.

In summary, at luminosities of $L_{opt}=10^{29}$  erg s$^{-1}$ Hz$^{-1}$ and
 less, a
variety of effects makes it hard to come to a clear conclusion, but above this
 luminosity
there is quite a strong case that the gross trend  of colour with luminosity is
 caused by
some kind of homologous scaling effect, caused for example by accretion rate
 varying by
several orders of magnitude. The range of colours at a fixed luminosity could
 correspond to
differing black hole masses, inclination effects, or to a relatively modest
 range of reddening.

\subsection{Absolute SED comparisons}
\label{comp-SEDs}

We can illustrate how the homologous scaling links high luminosity and low
 luminosity AGN
by directly
comparing the SEDs of 3C273 and M81. The optical-UV spectrum of M81 is well
 known for
being very
different from high luminosity AGN, showing $\alpha=2$ as opposed to
 $\alpha=0-1$. At
longer wavelengths,
it is not clear when small aperture nuclear measurements really are measuring
 quasar-like
emission as
opposed to normal stellar and stellar-related emission. Fig. 4 shows the effect
 of scaling
the SED of
3C273 with $X=0.03$. This produces a remarkably good fit to the mid-IR and
 optical data
on M81, with
the near-IR data looking like a stellar excess over this emission. (Note however
 that
both near-IR and mid-IR points are essentially upper limits to the true nuclear
 emission.)
The simplest bet for the mystery factor $X$ is of course
accretion rate. The implication would be that the nucleus of M81 is essentially
 the same
as 3C273 but
with accretion rate lower by a factor $(1/0.03)^4$ , in other words by roughly a
 factor of
a million.

The fact that this exercise gives roughly the right answer must be important.
 However the
impressively good fit should be taken with a pinch of salt, for several reasons.
 (i) The figure
shows the soft X-ray emission from 3C273 matching on to the optical-UV
emission for M81, but the soft X-ray emission is unlikely to be made up of
 multiple
black-bodies, and so shouldn't in itself show the homologous scaling. However,
 as it hangs off
the peak of the SED, if the optical-UV emission scales, the soft-X-ray emission
 will fall roughly
into
place. (This does however suggest that the soft X-ray is somehow physically 
 connected to the
blackbody
disc emission.)  (ii) Figure 4 also shows the SED of
NGC~5548 and an attempt to match a scaled version of 3C273 with $X=0.21$. This
 is less impressive
than the
fit with M81 but still plausible. 
The comparison is complicated by the well known variability of NGC~5548. Figure
 4 shows 
low, mean and high states from the extensive NGC~5548 AGN Watch literature. (All
 of these 
SEDs have been corrected for starlight.) The component which scales is
 presumably the 
non-varying part. (iii) The black holes in 3C273 and M81 may have 
different masses. Paltani and Turler (2003) estimate a mass of 
 $10^9 M_\odot$ for 3C273, and Devereux \et (2003) give $7\times 10^7 M_\odot$ 
for M81. Of course this difference of one order of magnitude is dwarfed by the
 possible six orders
of magnitude in accretion rate. (iv) Finally, scaling the SED of H1946+786
 doesn't fit  nearly so 
well.
For simplicity, I don't  show this. 

Overall, these tests support the idea that homologous scaling 
plays an important part in AGN SEDs,
but also show clearly that it will not explain all the facts.

%%%%%%%%%%%%%%%%%%%%%%%%%%%%%%%%%%%%%%%%%%%%%%%%%%%%%%%
% FIGURE-4 : ABSOLUTE SED SHIFTS
%
\begin{figure}
\includegraphics[width=60mm, angle=270]{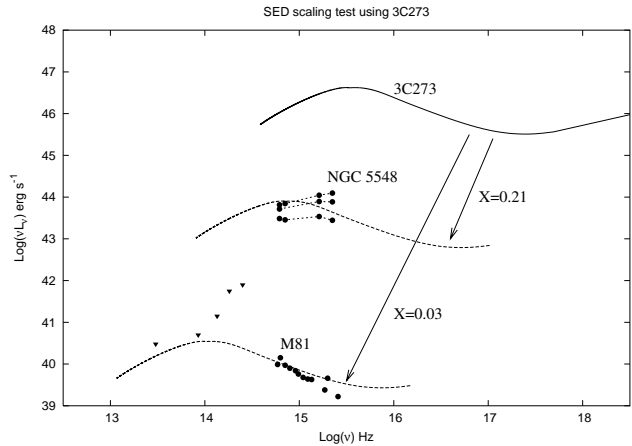}
\caption{Test of the hypothesis that the SEDs of NGC~5548 and M81 are 
homologously shifted versions of the SED of 3C273. }
\end{figure}
%
%%%%%%%%%%%%%%%%%%%%%%%%%%%%%%%%%%%%%%%%%%%%%%%%%%%%%%%

\section{Comparison with AGN variability data}  
\label{vblty}

There is a second well known colour-luminosity effect in AGN. Individual objects
 vary in 
luminosity by a factor of several, and become bluer as they become brighter. The
 best 
studied example is the Seyfert galaxy NGC 5548 (Clavel \et 1991; Peterson \et
 1991; Krolik \et 1991). 
A well 
known puzzle in this 
variability behaviour is that variations occur more or less simultaneously at
 different 
wavelengths, whereas viscous effects should produce delays on the order of
 hundreds 
to thousands  of 
years, and even instabilities should travel at sound speed, producing delays of
 the order 
of years. 
This led to the suggestion that AGN optical-UV emission actually arises from a 
passive disc heated by the central X-ray source or the inner disc (Krolik \et
 1991) in 
which case variations are transmitted at light speed. For a fixed disc geometry
 such 
driving variations in the central source heating should then produce
 colour-luminosity 
changes that fit the predicted locus from homologous scaling. 

Fig 2 shows the variability range for NGC 5548. Here I have taken the optical
 high and low 
states from Paper II of the AGN Watch series on NGC~5548 (Peterson \et 1991), at
 JD2447645 
and JD2447574 respectively. The optical spectral index $\alpha_{opt}$ was
 calculated using 
the ratio of fluxes at $4300 \rm\AA$ and $5800 \rm\AA$,  and $L_{opt}$
 calculated using the flux 
at $4300 \rm\AA$. The data in Peterson \et are reduced to a standard AGN Watch 
aperture. Starlight in the standard aperture was estimated by Romanishin \et
 (1995) at a 
wavelength of $5100 \rm\AA$. I extrapolated to other wavelengths using the
 standard bulge 
starlight colours of Ward \et (1987). It is clear that in the optical range, the
 starlight 
contribution through typical apertures in NGC~5548 is quite substantial, which
 in itself
can lead to the object being bluer when brighter. However, even for the
starlight corrected data, the change in colour with brightness is very steep, 
 and is 
clearly inconsistent with our predicted multi-temperature black-body track. In
 the UV, where 
the effect of starlight is negligible, the change of colour with brightness is
 also steep.
If then the emission from NGC~5548 originates from an extended multi-temperature
 blackbody, 
it is not the case that all the parts of that extended body vary together, as
 one would 
expect with the X-ray
source reprocessing model. 
(A similar conclusion was drawn by Berkeley, Kazanas, and Ozik (2000) in their 
attempt to model the simultaneous X-ray and UV monitoring data for  NGC~7469.)

The variability of NGC~5548 can in fact be 
quite well explained by a simple mixing model, with a fixed cool component and a
 varying
hot component. This is an appealing possibility because 
it may largely remove the lack-of-delay problem - delays relate only to the
 variable component, 
which could for example come from just the inner regions of the accretion disc.
Fig. 5 shows the UV colour-brightness effect 
using the data of Clavel \et (1991) at $1337 \rm\AA$ and $1813 \rm\AA$ 
downloaded from the AGN  Watch website. I modelled
these data using two components.  The first is constant in both colour and flux,
 and the 
second has fixed colour but variable normalisation. The difference between any
 two epochs
is due to only to the variable component, thus giving the fixed colour of the
 variable 
component. I estimated this by averaging the five lowest and five highest data
 points to
get $F_{1337}/F_{1813}=1.60$ corresponding to $\alpha=-0.46$ for the variable
 component. This
leaves one variable quantity (flux of variable component) and two free model
 parameters (flux
and colour of constant component). I adjusted the two free parameters by hand to
 fit the 
data. The result is an excellent fit to both the slope and the curvature of the
 effect. 
I have not gone beyond this to a formal best fit, as the model is only
 illustrative. It is far 
from clear what an apppropriate physical model might be, as discussed in section
 5.

%%%%%%%%%%%%%%%%%%%%%%%%%%%%%%%%%%%%%%%%%%%%%%%%%%%%%%%
% FIGURE-5 : N5548 MIXING MODEL
%
\begin{figure}
\includegraphics[width=60mm, angle=270]{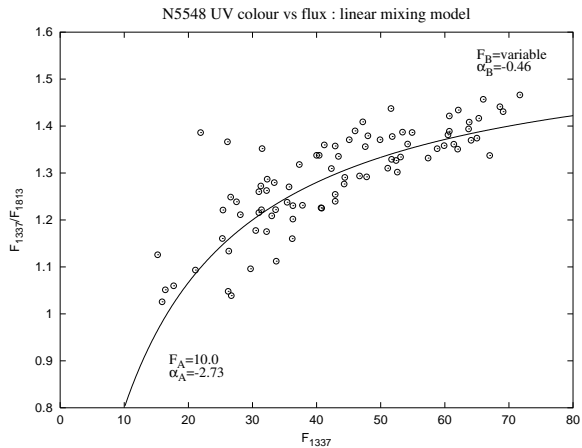}
\caption{Colour-luminosity effect in the UV emission from NGC~5548. Data from 
Clavel et al 1991, taken from the AGN Watch web site. The curve shown is the
 prediction of
a simple mixing model, with a constant red component and a variable blue
 component, 
as explained in the text.}
\end{figure}

%
%%%%%%%%%%%%%%%%%%%%%%%%%%%%%%%%%%%%%%%%%%%%%%%%%%%%%%%

\section{Discussion}
\label{discuss} 

The arguments presented here do not confront specific accretion disc models, but
 support 
the generic idea that AGN SEDs are dominated by an {\em extended object} 
radiating as the {\it sum of
blackbodies}. The recent results of Rokaki \et (2003) on aspect dependence of
 superluminal 
sources support the generic idea of a {\em flat} emitter, and those of Kishimoto
 \et 
(2003,2004) show the polarised Balmer edge absorption expected from an
 atmosphere above 
such a flat object.  These results taken together strongly support the simple
 idea 
of accretion discs. Beyond this I have argued that most of the trend seen in the
  $\alpha - L$ diagram
is due to differences in accretion rate, not differences in black hole mass. We
 are left with the 
challenge of explaining the variability seen. Beyond these simple statements,
 there are still 
significant problems and further work needed.

{\it Is the SED peak really moving ?} The argument made in this paper rests
 mainly on the general
trend of colour versus luminosity for the population of AGN, rather than
 detailed analysis of SEDs.
If homologous scaling applies to the whole SED, then the peak of the SED should
 be shifting 
in frequency through
the easily observable optical region in the lower  end of the Seyfert galaxy
 range. With reasonably 
complete SEDs for a range of objects this should be a testable hypothesis, and a
 good target for 
future work. However we can expect some complications that may muddy this simple
 test. Starlight, 
small 
blue bump, and reddening will all need to be accounted for; and  it is only at
 wavelengths longward 
of the 
peak that we are reasonably confident that scaling  will apply. The inner
regions may not be simple blackbodies; heating on the spot may break down; and
 the central
variable component may not scale.

{\it Can the $\alpha - L$ trend be explained by other effects ?} Starlight
 contamination,
 variability, and reddening are all clearly important. Is it possible that one
 or more of these
effects themselves
vary slowly with luminosity, producing the effect we see ?  This is most
 obviously a worry
for starlight contamination. I have argued in section 3.1 that this effect is
 only
important in the lower half of Fig. 2, and so does not explain the trend at
 higher luminosity. What
about
reddening ? It has been previously
suggested by Puchnarewicz \et (1996) and Gaskell \et (2004) that variation of
 reddening with
luminosity is precisely  what causes the colour-luminosity trends we have been
 discussing
here. This is indeed a tenable alternative hypothesis. The main argument
against it is the ``fine tuning'' worry - one has to postulate a reddening
screen, and have its thickness vary very slowly with luminosity in just the
 right manner,
for an unknown reason, whereas the tracks we have predicted in Fig.2 follow from
 the
chosen template SEDs with no adjustable parameters. An attractive possibility is
 that the
trend shown by the lower envelope of the points corresponds to the homologous
 scaling effect,
whereas the observed vertical spread of the points is due to a range of
 reddening at each
luminosity. Rather than the {\em minimum} reddening varying with luminosity,
 what probably
alters is the {\em probability of large reddening}. Finally another possible
 explanation of the colour-luminosity effect is that the optical-UV continuum
may be made of two components - for example, an accretion disc and an underlying
 power law, with the relative strengths of these varying with luminosity - the
 so-called "fading bump" idea. This has been discussed by a number of authors
(e.g. Zheng and Malkan 1993; McDowell \et 1989; Carleton \et 1987; Ward \et
1987).  As with the reddening trend model however, there is no obvious natural
reason to produce just the right fading of blue  bump strength with luminosity.

{\it Can we separate mass and accretion rate effects ? } AGN range in luminosity
 over six orders
of magnitude. What causes this range ? Simple considerations, and the toy
 modelling in Fig. 3, suggest
that changes in accretion rate will produce homologous scaling, whereas changes
 in black hole mass
will not. The agreement of the scaling prediction with the data of Figs. 2 and 4
 therefore suggests
that
accretion rate is a significant and perhaps even dominant factor in determining
 luminosity.
There is a large vertical spread in Fig. 2 which could be due
to differences in black hole mass. However there are several
other likely causes of the vertical spread - reddening, variability, and
 starlight. Possibly
understanding
these better would restrict the range of allowed black hole mass considerably.
 An obvious programme of
work is to produce $\alpha - L$ tracks for changes in black hole mass, to
 compare and contrast with
the
homologous scaling tracks. Unlike the accretion rate effect, the result would be
 dependent on the
detailed physical model, and in particular the correct $T(R)$ formula. The point
 of this paper is to
examine the most model independent factors possible, so I don't attempt this
 here.

Other available evidence suggests that the range in AGN luminosity is due in
roughly equal measure to black hole mass and accretion rate. The first type of
evidence is from accretion disc model fitting to AGN SEDs. For example, Fig. 2
from Szuszkiewicz, Malkan, and Abramowicz (1996) compiles results from a
variety of papers and shows a range of about $10^3$ in fitted values for both
mass and accretion rate. The second kind of evidence concerns dynamical mass
 estimates. Several tens of
AGN now have fairly reliable mass measurements from reverberation studies
(Kaspi \et 2000; Peterson \et 2004). These studies find that Seyfert galaxies
 and quasars
have masses in the range  $10^{7-9} M_\odot$ whilst covering four decades in
 luminosity.
Kaspi \et quantify the relationship between mass and luminosity.
Using three related methods they
find that $M\propto L^\beta$ where $\beta = 0.16$, $0.40$, and $0.55$.
Scott \et (2004) find a large scatter in far-UV slopes in a recent FUSE study of
 100 AGN, and
attribute
this
to AGN having a large range in both mass and accretion rate.
Amongst very low luminosity "dwarf Seyferts", there are few reliable mass
 estimates.
M81 has a mass of the same order as luminous AGN (Devereux \et 2003), but
 NGC~4395
is thought to have a very small mass, of the order $10^5 M_\odot$ (Lira \et
 1999;
Kraemer \et 1999; Shih \et 2003). However including this object
also extends the luminosity range even further, so we would  be talking
about four orders of magnitude of
mass compared to eight orders of magnitude in luminosity. NGC~4395 is also quite
 blue despite having
a very low luminosity, going against the trend of Fig. 2, as one might expect
 with a very small mass.

{\it Can accretion disc models explain the observed variability ?} The nature of
 the optical-UV
variability remains the most worrying and puzzling feature of AGN. Its steep
 colour dependence
and simultaneity at different wavelengths cannot be explained by standard
 accretion disc models
(Krolik \et 1991)
The usual explanation accepted for a decade was heating of a
passive disc by the central X-ray source,
but this model can now be rejected, firstly by the lack of the expected
UV -- X-ray variability correlation
(e.g. Nandra \et 1998; Berkeley, Kazanas, and Ozik 2000), and now by the lack of
 the expected
colour-luminosity scaling behaviour.
In section 5, I demonstrated that the
observed colour variations can be quite well modelled by mixing a constant red
 component and
a variable blue component. Can this be explained in the context of accretion
 discs ?
It seems reasonable that in fact the inner annuli of an accretion disc
may vary on a short timescale while the outer annuli remain much the same.
 Krolik \et (1991)
attempt to construct the power spectrum of variability in this way by allowing
 the frequency of
oscillations from different rings to vary with the orbital timescale of each
 ring, so that
optical variability for example mostly varies on a timescale of months, whereas
 UV variability
is on a timescale of days, with the optical tail of emission from the hot inner
 rings producing a
small amplitude of variability on such timescales.
However, as Krolik \et explain in their section 4.5.4, even within a small range
 of inner radii,
significant delays between different wavelengths are expected for fluctuations
 transmitting at
sound speed or even at orbital speed. Their proposed solution is hard photons
 from the inner disc
being reprocessed in the outer disc, producing co-ordinated fluctuations.
 However, we are then
back to square one, as such fluctuations due to changes in heating rate should
 follow
the locus discussed in this paper. One more alternative
is to abandon the accretion disc paradigm somewhere in the middle
regions, so that we can mix a static accretion disc with a central source of
 unknown nature
that is variable and has a broad SED.

A final possibility that has sometimes been mentioned
(e.g. Gaskell \et 2004) is that the optical-UV variability is
not intrinsic, but due to a changing reddening screen. This certainly gives
 roughly the
right slope in the colour-luminosity plane, but a cloud crossing the nucleus
 would occult
different parts of the disc at different times, giving wavelength dependent
 delays. A model of
this type would therefore need to involve radial changes in extinction, perhaps
 as a result of
changing
amounts of dust condensing in an outflowing wind.

\section{Conclusions}
\label{conclude}

Multi-temperature black-bodies should show a characteristic scaling behaviour
 when all
regions of the body change temperature by the same factor. For black hole
 accretion disc
models, we expect such a scaling to apply for changes of accretion rate or
external heating rate, but not  for changes of black hole mass.
The trend in optical colour for AGN over a very wide range in luminosity seems
 to be well 
fitted by such homologous scaling of the AGN SED. Several other effects
 significantly
affect AGN colours - starlight contamination, reddening, and variability.
 Separating
all these effects is extremely difficult. Nonetheless, the trends over a very
 wide
range in luminosity seem clear. The scaling effect seen applies only on average,
 amongst the 
population of AGN. From one typical object to another, other effects - black
 hole
mass, variability, starlight, and reddening, are much more likely to cause
 differences. Within
a single object, changes of colour with luminosity are not consistent with the 
expected scaling at all - during variability it is clearly not the case that all
 the
parts of the object vary together, as one would expect in the X-ray reprocessing
 model.
Instead, variability seems consistent with mixing a static component, which
 could be
identified with the accretion disc, with a variable
blue component, whose nature is unknown.

Overall a  number of considerations strongly support the general idea of
 accretion 
discs in AGN. The glaring exception is the optical-UV variability, and
 especially
its simultaneity at different wavelengths, which remains puzzling.

\section{Acknowledgements}  \label{acknow}

Since having the basic idea I have been not quite writing this paper up 
for about five years, despite brandishing it at several conferences.
My apologies to the various colleagues I have beeen promising it to, and
my thanks to a wide range of people for useful discussions, but especially
to Omar Almaini, Martin Elvis,  Makoto Kishimoto, Julian Krolik,  and 
Andrew King. 

%%%%%%%%%%%%%%%%%%%%%%%%%%%%%%%%%%%%%%%%%%%%%%%%%%%%%%%%%%%%%%%%%%%%%%%%
%    REFERENCES
%%%%%%%%%%%%%%%%%%%%%%%%%%%%%%%%%%%%%%%%%%%%%%%%%%%%%%%%%%%%%%%%%%%%%%%%%

\section{REFERENCES} \label{refs}

\noindent Berkley, A.J., Kazanas, D., Ozik, J., 2000,  ApJ, 535, 712
\smallskip

\noindent Carleton, N.P., Elvis, M., Fabbiano, G., Willner, S.P., Lawrence, A.,
Ward, M., 1991, ApJ, 318, 595
\smallskip

\noindent Clavel, J., Reichert, G.A., Alloin, D., Crenshaw, D.M., Kriss, G.,
 Krolik, J.H.,
Malkan, M.A., Netzer, H., Peterson, B.M., Wamsteker, W., and 47 coauthors, 1991,
ApJ, 366, 64
\smallskip

\noindent Devereux, N., Holland, F., Tsvetanov, Z., 2003, AJ, 125, 1226
\smallskip

%\noindent Elvis, M., Lawrence, A., 1985, in {\it Astrophysics of active
 galaxies
%and quasi-stellar objects}, ed. Milller, J., University Science Books, Mill
 Valley,
%CA, USA
%\smallskip

\noindent Elvis. M., Wilkes, B.J., McDowell, J.C., Green, R.F., Bechtold, J.,
 Willner, S.P.,
Oey, M.S., Palomski, E., Cutri, R., 1994, ApJS, 95, 1
\smallskip

\noindent Frank, J., King, A., Raine, D.J., 2000, {\it Accretion Power in
 Astrophysics},
3rd edition, Cambridge University Press
\smallskip

\noindent Gaskell, C.G., Goosman, R.W., Antonucci, R.R.J., Whysong, D.H, 2004,
ApJ in press (astro-ph/0309595)
\smallskip

\noindent Ho, L.C., Filippenko, A.V., Sargent, W.L.W., 1996, ApJ, 462, 183
\smallskip

\noindent Kishimoto, M., Antonucci, R., Blaes, O., 2004, MNRAS, 345, 253
\smallskip

\noindent Kishimoto, M., Antonucci, R., Boisson, C., Blaes, O., 2004, MNRAS
 submitted
\smallskip

\noindent Kraemer, S.B., Ho, L.C., Crenshaw, D.M., Shields, J.C., Filippenko,
 A.V., 1999,
ApJ, 520, 664
\smallskip

\noindent Kriss, G.A., Davidsen, A.F., Zheng, W., Lee, G., 1999, ApJ, 527, 683
\smallskip

\noindent Krolik, J. H., Horne, K., Kallman, T.R., Malkan, M.A., Edelson, R.A.,
 Kriss, G.A.,
1991, ApJ, 371, 541
\smallskip

\noindent Krolik, J.H., 1999, {\it Active Galactic Nuclei}, Princeton University
 Press
\smallskip

\noindent Kuhn, O., Bechtold, J., Cutri, R., Elvis, M., Rieke, M., 1995, ApJ,
 438, 643.
\smallskip

\noindent Lira, P., Lawrence, A., O'Brien, P., Johnson, R.A., Terlevich, R.,
 Bannister, N.,
1999, MNRAS, 305, 109
\smallskip

\noindent Malkan, M.A., 1991, in {\it Accretion Disk Models for Active Galactic
Nuclei}, ed. Bertout, C., Collin-Souffrin, S., and Lasota, J.P.,
Editions Frontieres, Gif-sur-Yvette
\smallskip

\noindent Mushotzky, R.F., Wandel, A., 1989, ApJ 339, 674
\smallskip

\noindent McDowell, J.C., Elvis, M., Wilkes, B.J., Willner, S.P., Oey, M.S.,
Polomski, E., Bechtold, J., Green, R.F., 1989, ApJLett 345, L13
\smallskip

\noindent Nandra, K., Clavel, J., Edelson, R.A., George, I.M., Malkan, M.A.,
 Mushotzky, R.F.,
Peterson, B.M., Turner, T.J., 1998, ApJ, 505, 594.
\smallskip

\noindent Paltani, S., Turler, M., 2003, ApJ, 583, 659
\smallskip

\noindent Peterson, B. M., Balonek, T.J., Barker, E.S., Bechtold, J., Bertram,
 R., 
Bochkarev, N.G., Bolte, M.J., Bond,  D., Boroson, T.A., Carini, M.T., 
and 54 coauthors, 1991, ApJ, 368, 119
\smallskip

\noindent Peterson, B.M., Ferrarese, L., Gilbert, K.M., Kaspi, S., Malkan, M.A.,
Maoz, D., Merritt, D., Netzer, H., Onken, C.A., Pogge, R.W., 
Vestergaard, M., Wandel, A, 2004, preprint (astro-ph/0407299)   
\smallskip

\noindent Puchnarewicz, E.M., Mason, K.O., Romero-Colmenero, E., Carrera, F.J., 
Hasinger, G., McMahon, R., Mittaz, J.P.D., Page, M.J., 
Carballo, R., 1996, ApJ, 281, 1243
\smallskip

\noindent Rokaki, E., Lawrence, A., Economou, F., Mastichiadis, A, 2003, MNRAS,
 340, 1298.
\smallskip

\noindent Rokaki, E., Magnan, C., 1992, A\&A, 261, 41
\smallskip

\noindent Romanishin, W. Balonek, T.J., Ciardullo, R., Miller, H.R., Peterson,
 B.M.,
Sadun, A.C., Stirpe, G.M., Takagishi, K., Taylor, B.W., 
Zitelli, V., 1995, ApJ, 455, 516
\smallskip

\noindent Scott, J.E., Kriss, G.A., Brotherton, M., Green, R.F., Hutchings, J.,
 Shull, J.M.,
Zheng, W., 2004, ApJ in press (astroph/0407203)

\noindent Shih, D.C., Iwasawa, K., Fabian, A.C., 2003, MNRAS, 341, 973
\smallskip

\noindent Sun, W.-H., Malkan, M.A., 1987, ApJ, 346, 68.
\smallskip

\noindent Szuszkiewicz, E., Malkan, M.A., Abramowicz, M.A., 1987, ApJ, 458, 474.
\smallskip

\noindent Ward, M., Elvis, M., Fabbiano, G., Carleton, N.P., Willner, S.P., 
Lawrence, A., 1987, ApJ, 315, 74
\smallskip

\noindent Zheng, W., Kriss, G.A., Telfer, R.C., Grimes, J.P., 
Davidsen, A.F., 1997, ApJ,  475, 469 
\smallskip

\noindent Zheng, W., Malkan, M.A., 1993, ApJ, 415, 517.
\smallskip

%+++++++++ CLOSE OUT +++++++++++++++++++++++++++++++
\label{lastpage}

\end{document}